\documentclass[epj]{svjour}
\usepackage[utf8]{inputenc}
\usepackage{graphicx,booktabs,amsmath,amssymb,url,soul,color}
\begin{document}
\title{Spatial firm competition in two dimensions with linear transportation costs: simulations and analytical results}
\titlerunning{Spatial firm competition in two dimensions with linear transportation costs}
\author{Alan Roncoroni and Matúš Medo\thanks{matus.medo@unifr.ch}}
\date{\today}
\institute{Department of Physics, Chemin du Musée 3, 1700 Fribourg, Switzerland}
\abstract{Models of spatial firm competition assume that customers are distributed in space and transportation costs are associated with their purchases of products from a small number of firms that are also placed at definite locations. It has been long known that the competition equilibrium is not guaranteed to exist if the most straightforward linear transportation costs are assumed. We show by simulations and also analytically that if  periodic boundary conditions in two dimensions are assumed, the equilibrium exists for a pair of firms at any distance. When a larger number of firms is considered, we find that their total equilibrium profit is inversely proportional to the square root of the number of firms. We end with a numerical investigation of the system's behavior for a general transportation cost exponent.}
\maketitle

\section{Introduction}
The problem of firm competition in an economy has been studied extensively in the past, leading to the classical concepts such as perfect competition, price equilibrium, and oligopoly~\cite{varian2010intermediate}. Among the models of competition, two classical models stand up: the Cournot competition where the participating companies decide the produced amount and Bertrand competition where the companies decide the product price. Depending on circumstances, the former or the latter may be more appropriate to model a given situation. The rather extreme assumptions made by the Bertrand model lead to a so-called Bertrand paradox: in equilibrium, all participating firms earn zero profits~\cite{bertrand1883theorie,edgeworth1897teoria}. This is because the firms have an incentive to decrease the price and thus attract all the consumers in the market, and the spiral of price decrease does not stop until the product marginal cost equals the product price for each firm, and thus zero profit is made by all. These assumptions can be relaxed by considering product differentiation~\cite{shaked1982relaxing} and non-price competition~\cite{clay2002retail}, production capacities~\cite{kreps1983quantity}, consumer search costs~\cite{anderson1999pricing}, and---the key point of interest on this paper---transportation costs~\cite{hotelling1929stability}.

In the real world, market activities usually occur at different points in space which makes it important to include transportation costs in our considerations of economics. Consumer-side transportation costs have been first introduced in a classical paper on firm competition by Hotelling where the author considers the case of consumers distributed uniformly on a line of length $l$ and assumes that product price (often referred to as ``mill price'' in the literature) is augmented by the transportation cost which is proportional to the distance between the consumer and the firm~\cite{hotelling1929stability}. This simple setting makes it possible to study the price competition of firms and, assuming that the firms are free to choose their location, also the question of the optimal location. As a result, the paper has initiated an extensive line of research in spatial competition (see~\cite{eiselt1989competitive,biscaia2013models} for reviews).

However, 50 years after the original article was published, an important flaw has been discovered in its analysis: the price equilibrium does not exist when the two competing firms are close, unless one switches from linear to quadratic transportation costs~\cite{d1979hotelling}. It has been shown later that quadratic transportation costs are in fact the only one in  the family of power-law transportation cost functions for which the price equilibrium exists---all other powers share the flaw of linear transportation costs when the firms are located sufficiently close to each other~\cite{champsaur1988existence,tabuchi1994two}. Furthermore, the original conclusion that for the firms it is advantageous to be close to each other~\cite{hotelling1929stability} (so-called principle of minimum differentiation~\cite{boulding1966economic}) changes dramatically under quadratic transportation costs as it becomes advantageous for the firms to be as much apart as possible~\cite{d1979hotelling,hay1976sequential} (see~\cite{brenner2001determinants} for a survey on equilibrium existence and product differentiation). Note that one refers to product differentiation here because the firm's position in physical space can be interpreted as the product's position in the space of product properties; the transportation cost consequently becomes the additional cost attributed to a mismatch between the product's properties and the consumer's preferences~\cite{irmen1998competition}.

Despite their limited relevance to real economy, studies of the one-dimensional case (so-called ``Hotelling's beach'') prevail because of its mathematical tractability~\cite{smithies1941optimum,eaton1975principle,egli2007hotelling}. In some works, a circular market is considered (in the physics terminology, periodic boundary conditions in one dimension are assumed) which has profound influence on the existence of equilibrium and its nature~\cite{salop1979monopolistic,kats1995more}. Analyses of the two-dimensional~\cite{tabuchi1994two} and multi-dimensional~\cite{irmen1998competition} case exist but, motivated by the analytical studies of the one-dimensional case, assume quadratic transportation costs. By contrast, we use analytical and simulation techniques to study the Hotelling model in plane with periodic boundary conditions. We first show that with linear transportation costs, the Nash equilibrium of two firms exists and present analytical expressions for the equilibrium price and profit. The difference between the case with and without periodic boundary conditions is discussed in detail. We then study a situation where many firms compete and investigate how the equilibrium's properties depend on the number of firms. We consider here also the general case of transportation costs that grow as a power of distance; linear and quadratic transportations costs are special cases of the general case.

In our study, we assume almost exclusively periodic boundary conditions (PBCs) which are generally favorable for numerical simulations as they help to suppress the finite-size effects. With two competing firms, PBCs can be interpreted as the influence of ``external'' firms beyond the simulated region's boundary. When there are many competing firms, PBCs assure that apart from statistical fluctuations in firm positions, no firm is in a privileged location. Without PBCs, firms in the middle of the studied region are completely surrounded by competing firms (they effectively experience PBCs) and their location is thus considerably less advantageous than that of a firm with no competitors between the firm and a boundary. Finally, we will show that PBCs are crucial in making the equilibrium of firm competition stable and analytically tractable.

\section{Model}
We assume that the customers are uniformly distributed in the unit square $[0, 1]\times[0,1]$. In the discrete version of the model, which we employ for simulations, the customers are labeled with a pair of indices. The coordinates of customer $(i, j)$ are $(x_i, y_j)$ where $x_i = (i - 0.5) / N$, $y_j = (j - 0.5) / N$ and $i, j = 1,\dots, N$; there are thus $N^2$ customers in total. Assuming that there are $m$ firms in the unit square, we label their coordinates as $(X_k, Y_k)$ where $k=1,\dots, m$. Since we consider a domain of unit area, the number of customers $N^2$ and the number of firms $m$ are equivalent to the customer and firm density, respectively.

The offered product price of firm $k$ is $p_k$. The effective cost $E_k(i, j)$ that customer $(i, j)$ has to pay for the product of firm $k$ consists of the product price and the transportation costs. Denoting the transportation cost over a unit distance as $r$ and the exponent of the distance dependence as $\gamma$, we write
\begin{equation}
\label{eff_cost}
E_k(i, j) = p_k + r \big[\Delta(x_i, X_k)^2 + \Delta(y_j, Y_k)^2\big]^{\gamma / 2}.
\end{equation}
A multiplying factor of two corresponding to the travel to the location of firm $k$ \emph{and back} is assumed to be included in the multiplier $r$ for simplicity. Note that while $\gamma = 1$ results in transportation costs that depend linearly on the customer-store distance, $\gamma = 2$ reproduces the much-studied case of quadratic transportation costs.

\begin{figure}
\centering
\includegraphics[scale = 0.7]{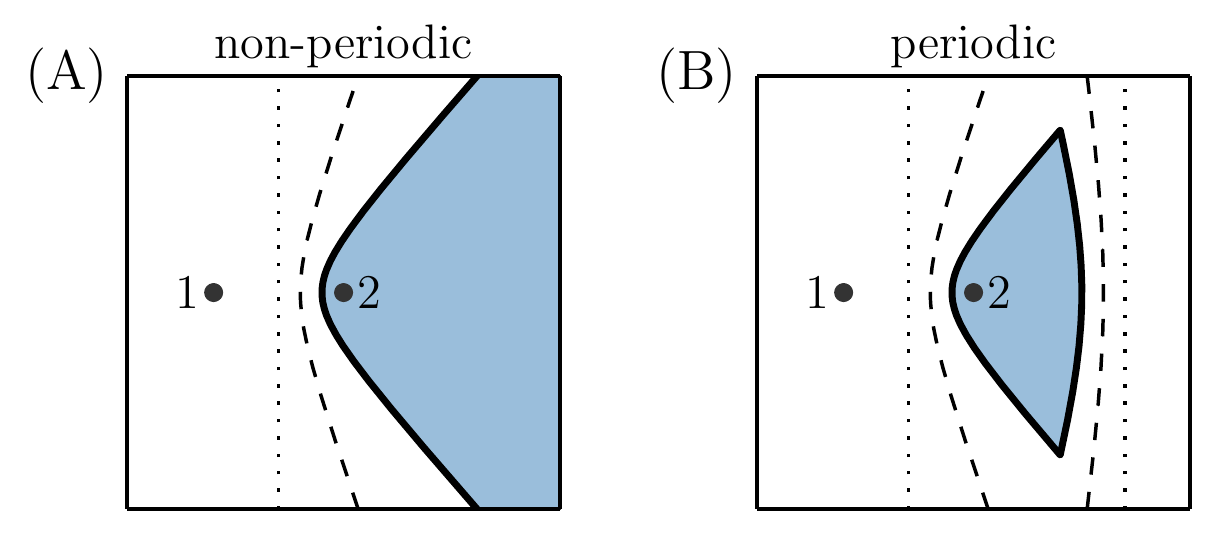}
\caption{By setting their product prices, the firms divide the customers among themselves. The main example here (solid lines and shaded regions) is of firm 1 at $(0.2, 0.5)$ with price $0.8$ and firm 2 at $(0.5, 0.5)$ with price $1$ (linear transportation costs with $r=1$). For comparison, we show also the results when the product price of firm 1 is $0.9$ (dashed line) and $1.0$ (dotted line). Panels A and B show the case without and with periodic boundary conditions, respectively.}
\label{fig:illustration}
\end{figure}

The coordinate difference $\Delta(\cdot, \cdot)$ in Eq.~(\ref{eff_cost}) can be considered simply as the absolute value of the difference between the two coordinates; this corresponds to the model variant with non-periodic boundary conditions. The form of $\Delta(\cdot, \cdot)$ is different under periodic boundary conditions where it reads
\begin{equation}
\Delta(x_i, X_k) =
\begin{cases}
    \lvert x_i - X_k\rvert & \text{for } \lvert x_i - X_k\rvert\leq 1/2,\\
1 - \lvert x_i - X_k\rvert & \text{for } \lvert x_i - X_k\rvert > 1/2.
\end{cases}
\end{equation}
That is, the customer chooses the shorter of the two possible paths: either within the unit square or across the unit square's boundary. The form of $\Delta(y_j, Y_k)$ is analogous. Under periodic boundary conditions, the shortest path between $(0.9, 0.5)$ and $(0.2, 0.5)$ has the length of $0.3$ as opposed to $0.7$ when periodic boundary conditions are not considered. As illustrated in Figure~\ref{fig:illustration}, the choice of boundary conditions has profound consequences on firm competition when the number of firms is small.

In the discrete version of the model where $N^2$ individual customers are present, we assume that each of them chooses the firm $k$ that minimizes the effective cost. This mathematically corresponds to minimizing $E_k(i,j)$ given by Eq.~(\ref{eff_cost}) with respect to $k$. Note that, as typical for Bertrand-like models, there is no upper bound for the price that the customers are willing to pay~\cite{varian2010intermediate}. A monopolist firm would therefore earn an arbitrarily high profit in this setting. Denoting the number of customers who choose firm $k$ as $N_k$, the profit of firm $k$ is then $p_k N_k$ (we assume here for simplicity that the products are produced at zero cost; another view at this is that $p_k$ are ``excess'' prices beyond the product's production costs). To remove the profit dependence on $N$, it is advantageous to consider firm profit per customer $X_k = p_k N_k / N^2$. The discrete version is convenient for numerical simulations where customers minimize their effective prices; once firm positions and prices are known, it is straightforward to compute $N_k$ and $X_k$ for all firms. A continuous version of the model, which formally corresponds to the limit $N\to\infty$ of the discrete model, is convenient for an analytical treatment. In the continuous case, we divide the unit square into regions, each of whose includes all points for which a given firm minimizes the effective cost (we can again assume either periodic or non-periodic boundary conditions). Assuming that the region belonging to firm $k$ has surface $S_k$, the profit of firm is $X_k = p_k S_k$ which is a continuous analog of the previous form $X_k = p_k N_k$. The convenience of the continuous version lies in the fact that both $p_k$ and $S_k$ can be changed infinitesimally; we use this in Section~\ref{sec:Nash} to analytically solve the Nash equilibrium for two competing firms at an arbitrary distance.

\section{Simulations results}
To illustrate the emergence of a competition equilibrium in the case with periodic boundary conditions, we consider two firms at distance $d$. The initial product price, which we set to 0.3 in our simulations, turns out to be unimportant for the long-term behavior of the system. Simulations proceed in steps in which the firms alternate in their attempts to maximize their profit. For example, when firm 1 optimizes its profit, we maximize $X_1(p_1\vert p_2)$ with respect to $p_1$ assuming that $p_2$ is given and fixed. In simulations, we carry out 120 consecutive profit optimizations. To avoid the influence of the initial conditions, the first 80 price and profit values are excluded from the evaluation of results.

\begin{figure}
\centering
\includegraphics[scale = 0.7]{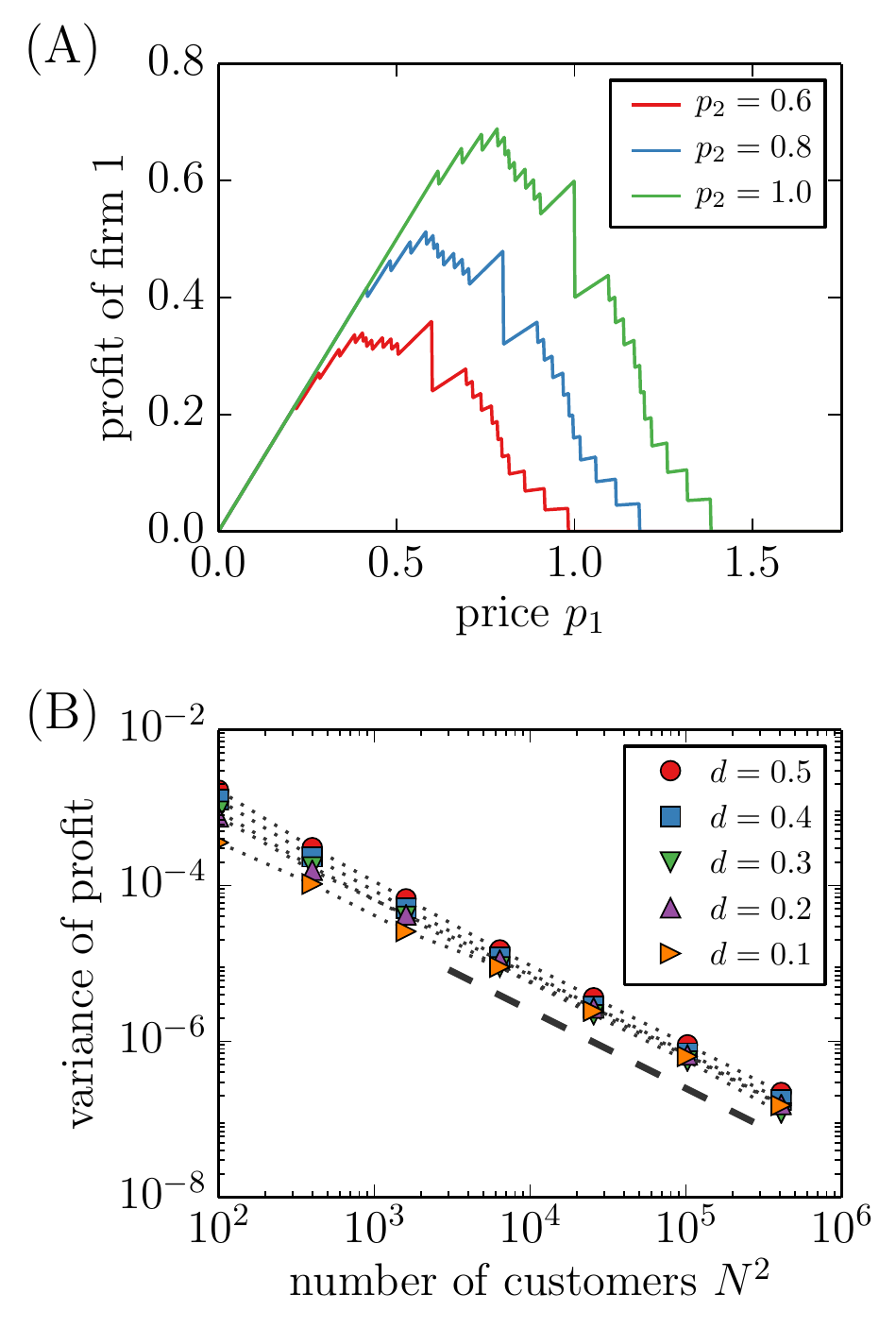}
\caption{Numerical investigation of the firm competition. (A) Profit of firm 1 per customer as a function of product price $p_1$ for various product prices $p_2$ of the competing firm at distance $d=0.5$. When $N$ is larger than $N=10$ assumed here, the profit profile becomes correspondingly smoother, yet it remains discontinuous. (B) The average variance of firm profit in consecutive optimization steps as a function of the system size. The indicative dashed line has slope $-1$.}
\label{fig:numerical}
\end{figure}

In simulations of a discrete system with $N^2$ customers, the profit of a firm is a discontinuous function of price: it grows linearly almost everywhere except for a finite set of points where one (or more) customers change from one firm to another (see Figure~\ref{fig:numerical}A for an illustration). Because of this discontinuous behavior, we do not use any of the standard maximization methods but simply evaluate the profit profit for 10,000 evenly-spaced price values in the range $[0,1]$ to find the optimal response of firm 1 to the price set by firm 2.\footnote{To prevent the emergence of periodic profit patterns, we used 100,000 evaluation points for $N=640$ (the last evaluation point in Figure~\ref{fig:numerical}B) and $d=0.5$.}
We find that for any finite value of $N$ and general initial conditions, prices and profits exhibit variations that do not vanish with simulation rounds. To show that the model actually leads to an equilibrium in the limit $N\to\infty$, Figure~\ref{fig:numerical}B shows that the variance of firm profit is proportional to $1/N^2$ and thus vanishes in the thermodynamic limit.

\section{Analytical results}
\label{sec:Nash}
To analytically study the competition equilibrium for two firms, we assume that the positions of firms 1 and 2 are $(0, 0.5)$ and $(d, 0.5)$, respectively, where $d\in(0, 0.5]$ (due to periodic boundary conditions, $d\in(0.5, 1)$ is equivalent with a corresponding smaller value $d'=1-d$). While the exact choice of firm position is important in the discrete case, in the continuous case with periodic boundary conditions, it is only the mutual distance of firms, $d$, what matters.

By choosing their prices $p_1$ and $p_2$, the firms divide the plane into two parts: in their effort to minimize the effective costs, customers in region 1 choose firm 1 over firm 2 and vice versa (see Figure~\ref{fig:illustration}A for an illustration). Boundaries of the regions are characterized by the effective costs of the two firms being equal along them. Points $(x_l, y_l)$ on the left boundary thus satisfy the condition
\begin{equation}
\label{left_boundary}
\begin{split}
r\sqrt{x_l^2 + (y_l - 0.5)^2} + p_1 =\\
=r\sqrt{(x_l - d)^2 + (y_l - 0.5)^2} + p_2
\end{split}
\end{equation}
and points $(x_r, y_r)$ on the right boundary satisfy
\begin{equation}
\label{right_boundary}
\begin{split}
r\sqrt{(x_r - 1)^2 + (y_r - 0.5)^2} + p_1 =\\
=r\sqrt{(x_r - d)^2 + (y_r - 0.5)^2} + p_2.
\end{split}
\end{equation}
Note that the right boundary is due to periodic boundary conditions; without them, only the left boundary exists.

Since the two firms are equivalent, the equilibrium price $p^*$ must be the same for both of them. The area of both region 1 and 2 is then equal to $1/2$. The two regions are then divided by parallel vertical boundaries: one at $x_l = d/2$ and the other at $x_r = (1 + d) / 2$ (this is true for any identical product prices $p_1=p_2$, not only for the Nash equilibrium of firm competition). To find the equilibrium price, we consider a small perturbation of the equilibrium by, say, firm 2 changing its price to $p^*+\Delta p$. The new profit of firm 2 is then
\begin{equation}
\begin{split}
X_2'(p^*+\Delta p, p^*) = (p^*+\Delta p)S_2(p^*+\Delta p, p^*) = \\
=(p^*+\Delta p)\int_0^1 \big[x_r(y) - x_l(y)\big]\,\mathrm{d}y
\end{split}
\end{equation}
where $x_r(y)$ and $x_l(y)$ are the corresponding left and right boundary of region 2 when product prices are $p^*$ and $p^*+\Delta p$, respectively.

\begin{figure}
\centering
\includegraphics[scale = 0.7]{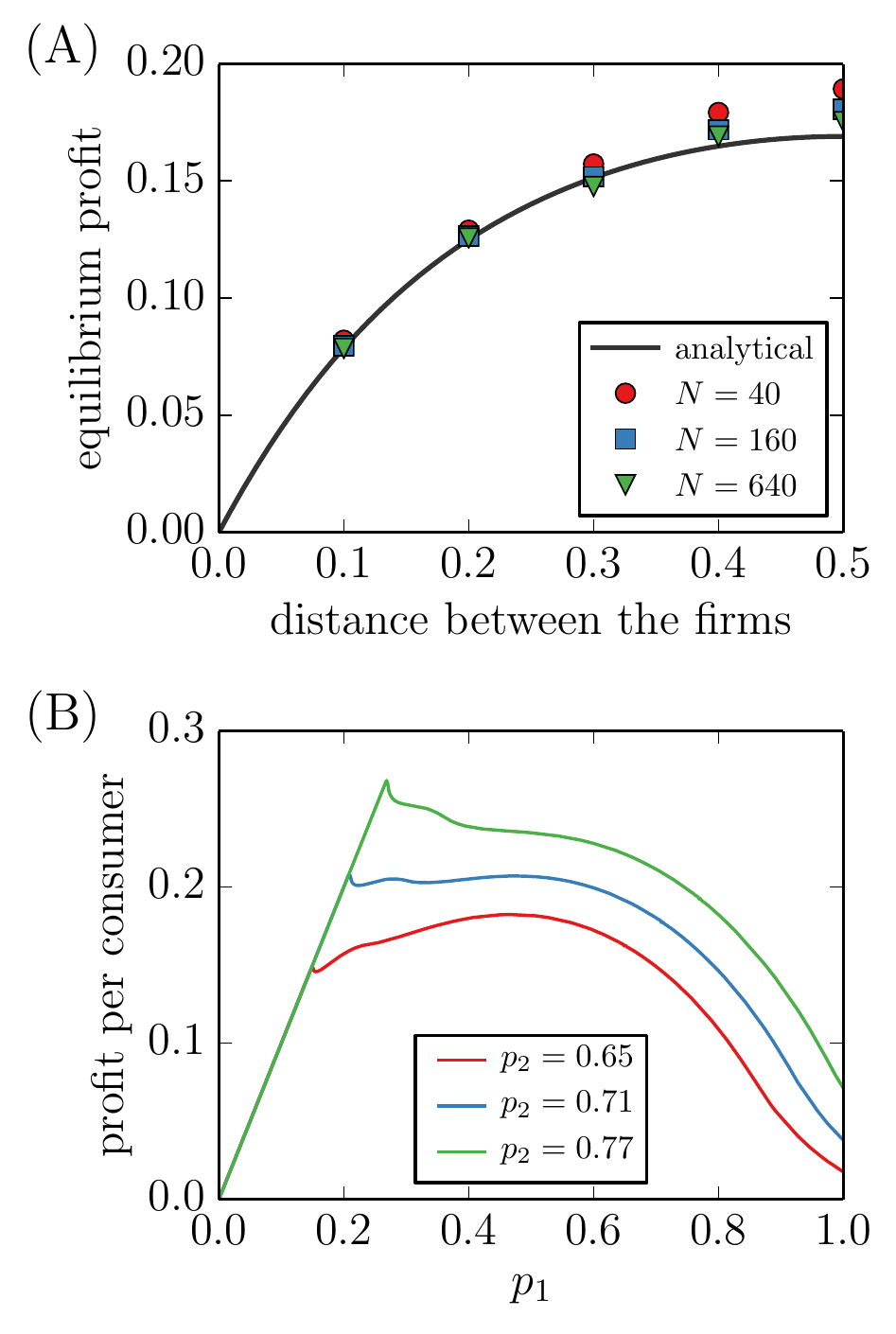}
\caption{(A) The average equilibrium profit per customer versus the distance between the two firms: a comparison between the analytical solution and numerical results obtained on systems with different size. (B) When periodic boundary conditions are not assumed, the profit of firm 1 can have multiple maxima (here for $p_2 = 0.71$) which leads to abrupt changes of price and prevent the equilibrium from emerging (firms 1 and 2 are here located at $[0, 0.5]$ and $[0.5, 0.5]$, respectively).}
\label{fig:Nash_profit}
\end{figure}

If $p^*$ is indeed the equilibrium price, the new profit of firm 2 must be the same as in equilibrium (up to $O(\Delta p^2)$). By doing the algebra, we eventually find the profit of each firm in the Nash equilibrium in the form
\begin{equation}
\label{nash_profit_analytical}
X^*(d, r) = \frac{(1-d)dr}{\Omega(d)}
\end{equation}
where
\begin{equation}
\begin{split}
\Omega(d) &= d\sqrt{(1 - d)^2 + 1} + (1 - d)\sqrt{d^2 + 1}+ \\
& + 3d(1 - d)^2\ln\left(1 - d\right) + 3d^2(1 - d)\ln d - \\
& - d(1 - d)^2\ln\left((1 - d)^2\left[\sqrt{(1 - d)^2 + 1}-1\right]\right) - \\
& - d^2(1 - d)\ln\left(d^2\left[\sqrt{d^2 + 1}-1\right]\right).
\end{split}
\end{equation}
In the derivation, we benefit from the fact that a small change of the price by one of the firms changes the original straight boundaries only infinitesimally. One can immediately note that the resulting Nash profit is directly proportional to the transportation cost rate $r$ and that it shows the expected symmetry $X^*(d, r) = X^*(1 - d, r)$ which is a direct consequence of assuming the periodic boundary conditions. Since the area of region 2 in equilibrium is $1/2$, the equilibrium price follows immediately from $X^*(d, r) = p^*(d, r) \times \tfrac12$. As shown in Figure~\ref{fig:Nash_profit}, the obtained analytical formula is in a good agreement with results of numerical simulations (although, convergence to the analytical result is rather slow).

\subsection{Equilibrium existence}
Similarly as shown in~\cite{d1979hotelling} for the one-dimensional case, the competition equilibrium does not exist when the periodic boundary conditions are not assumed. Instead, periodic patterns emerge where prices chosen by the firms slowly grow for a number of turns until it becomes profitable for the worse-located firm (\emph{i.e.}, the one closer to the unit square boundary) to substantially lower its price and thus attract all the customers. This can be seen in Figure~\ref{fig:Nash_profit}B where the profit-maximizing price $p_1$ is around 0.50 when $p_2=0.65$ and changes discontinuously to $0.20$ when $p_2\approx 0.71$. Note that for $p_2=0.71$ (and above), the maximum lies on the line $X(p_1)=p_1$ which means that it is indeed achieved by attracting all the customers at the expense of the other firm. It follows from Eq.~(\ref{left_boundary}) that the highest price $p_1$ that still attracts all the customers is $p_1' = p_2 - rd$ where $d$ is the distance between the firms. If firm 1 cannot achieve higher profit by setting price $p_1>p_1'$, where the lost of customers would be compensated by the increased price, $p_1'$ emerges as a local maximum. As shown in Figure~\ref{fig:Nash_profit}B, this occurs only when $p_2$ is sufficiently high; only then has firm 1 the incentive to increase its profit by dramatically reducing its price and thus attracting all the customers.

To understand the abrupt changes of product price quantitatively, assume that two firms located at $[0, 0.5]$ and $[d, 0.5]$ have reached the state $(p_1, p_2)$ where both $p_1$ and $p_2$ are local maxima of the corresponding firm's profit. The area claimed by the worse-positioned firm 1 is $S_1$. The current state $(p_1, p_2)$ is assumed to be locally optimal but it is stable only if firm 1 does not have the incentive to undercut firm 2 by lowering its price to attract all the customers. The profit achieved by doing so would be $p_1'\times 1$ as opposed to the current firm's profit $p_1 \times S_1$. It thus follows that the current state is stable only if
\begin{equation}
\label{stability}
p_1'<p_1 S_1.
\end{equation}
This condition is only fulfilled when the two firms are sufficiently apart. By taking, for example, $d=0.2$, one can find numerically that $p_1\approx0.12$, $p_2\approx0.24$, and $S_1\approx0.30$ (we assume $r=1$ here). Therefore $p_1'=0.04$ and the inequality above is violated because $p_1S_1\approx0.036$. The local maximum is thus not stable and an endless series of price adjustments by the two firms ensues instead of an equilibrium.

We return now to the case with periodic boundary conditions which is simpler to analyze because any local maximum must be necessarily symmetric. Starting in a local maximum $(p^*,p^*)$, it is not favorable for a firm to set price $p'=p^*-rd$ and thus attract all the customers if thus-achieved profit is smaller than the current one. The current area claimed by both firms is $1/2$. Equation~(\ref{stability}) thus takes the form $p^* - rd < p^*/2$ which implies $p^*<2rd$. One can easily verify that the equilibrium price $p^*=2X^*$ that follows from Eq.~(\ref{nash_profit_analytical}) satisfies the obtained stability inequality over the whole range $d\in[0,0.5]$ and the local maximum $(p^*,p^*)$ is therefore always stable.

To conclude, we write the stability condition again in the form
\begin{equation}
p_1 > \frac{p_1 + (p_2 - p_1) - rd}{S_1}.
\end{equation}
In the periodic case, $p_2=p_1$ and $S_1$ is increased by the fact that firm 1 can also attract customers from the region ``behind'' firm 2. In the non-periodic case, $p_2>p_1$ (the better-positioned firm can afford asking a higher price) and $S_1$ is smaller than in the periodic case. The corresponding stability inequality is therefore stronger and as the firms get closer to each other, the inequality is eventually violated and stability lost. These findings are in parallel with the study of the one-dimensional case presented in~\cite{kats1995more} where the introduction of periodic boundary conditions also restores the existence of an equilibrium in pure strategies.

\section{Competition of multiple firms}
As we mentioned in Introduction, the main reason for assuming periodic boundary conditions is the fact that in many situations, there are many firms competing in the market. To further investigate this case, we now study how the equilibrium profit changes with the number of firms in the unit square. It is clear that in the general case with $m$ firms, their precise mutual positions determine the equilibrium profit of each of them. The situation is therefore much more complex than in the case of two firms where their distance $d$ is the only variable. To refrain from unnecessary details, we assume that the firms are distributed in the plane at random with uniform probabilistic density. We aim to characterize the average equilibrium profit in such a situation with emphasis on the case of $m\gg 1$.

Even when the firms are placed at random, the equilibrium price of a particular firm is decided by the distances of several firms that surround it. In particular, the firms that lie in the adjacent regions of the Voronoi tessellation of the plane are the ones with which the studied firm has direct contact and competes for customers. The equilibrium prices of those firms are further decided by their neighboring competing firms, and so forth, and it is thus easy to see that the situation is not analytically approachable without making a further simplifying assumption. Since the strongest competition is with the closest neighboring firm (as shown in Figure~\ref{fig:Nash_profit}, equilibrium profit increases with distance between the firms), our simplifying assumption is that the equilibrium profit is decided solely by the distance of the closest firm. This effectively breaks the afore-described infinite chain of firm interactions and returns us to the two-firm case that we have studied above.

To compute the average distance of the closest firm, we use extreme statistics and compute the probability that the closest firm is at distance $D\in[R,R+\Delta R)$ ($\Delta R\to0$). This probability is composed of three factors: the probability that one firm is at distance $[R,R+\Delta R]$, the probability that the remaining $m-2$ firms are not closer than $R+\Delta R$, and $m-1$ which corresponds to the fact that any of the remaining $m-1$ firms can be the closest one. Taken together, we have
\begin{equation}
P(R\leq D<R+\Delta R) = 2\pi R\Delta R\varrho\,(1 - \pi R^2\varrho)^{m - 2} (m-1)
\end{equation}
which further simplifies if we plug in the uniform probability density $\varrho = 1$. Note that we neglect here the square geometry of the unit square and the periodic boundary conditions; we can do that because the closest firm is typically close enough to make these effects unimportant (especially when $m$ is large). Integration of $P(R\leq D<R+\Delta R)$ over $R$ from $0$ to $1/\sqrt{\pi}$ (the value at which the probability density decreases to zero) shows that this probability density is properly normalized. We finally compute the average distance of the closest firm as
\begin{equation}
\begin{split}
\overline{D} &= \int_0^{1/\sqrt{\pi}} R P(R)\,\mathrm{d}R = \frac{m-1}{2}\,\frac{\Gamma(m - 1)}{\Gamma(m + 1/2)} =\\
&= \frac{1 + O(1/m)}{2m^{1/2}}
\end{split}
\end{equation}
where we used the asymptotic form $\Gamma(x+n)/\Gamma(x)= x^n [1 + O(1/x)]$ for $x\to\infty$. This result agrees well with numerical simulations (figure not shown). Note that the scaling of $\overline{D}$ with $m$ is easy to obtain by arguing that $\overline{D}$ creates a region with area $\pi \overline{D}^2$ around each firm and the total number of those areas should roughly match the total area of the unit square. We can thus write $m\pi\overline{D}^2\approx1$ and therefore $\overline{D}\propto 1/\sqrt{m}$ which scales with $m$ as we derived above.

Eq.~(\ref{nash_profit_analytical}) can now be multiplied with two to yield the total profit of firms in the two-firm situation. By substituting the obtained $\overline{D}$ for $d$ in this result and further dividing with $m$, we obtain the profit per firm in the general situation with $m$ firms. Since the resulting form of the firm profit is very convoluted, we work out the leading contribution in the limit $m\to\infty$ which has the form
\begin{equation}
\label{number_of_firms}
X(m) = \frac{r}{m^{3/2}} + O(1/m^2).
\end{equation}
Looking back at the total profit of the competing firms, we get $X_{\text{total}}(m) = mX(m) = r / \sqrt{m}$ which tells us that the total profit of all involved firms decays as the number of firms grows.

\begin{figure}
\centering
\includegraphics[scale = 0.7]{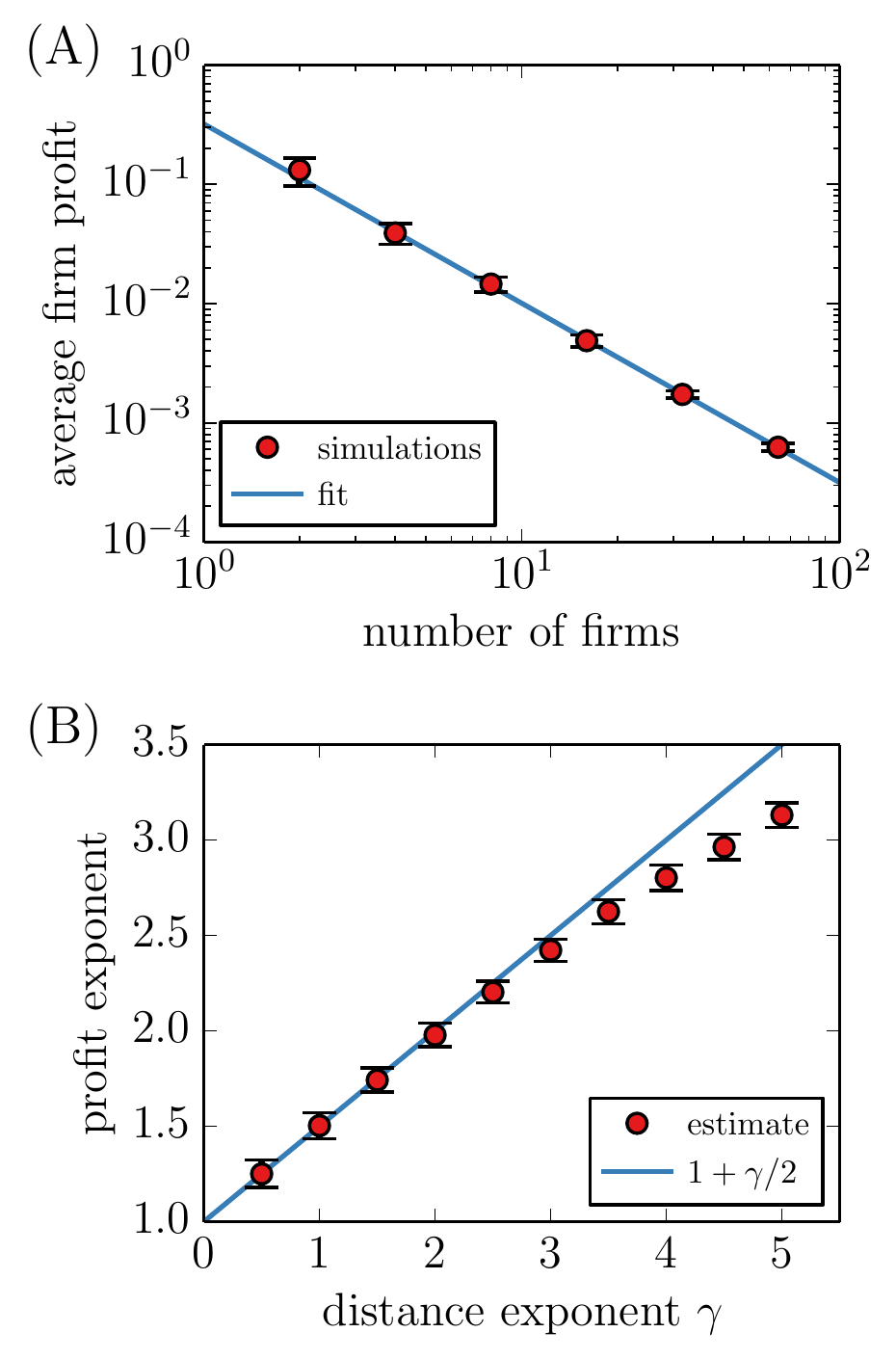}
\caption{(A) The dependence of the equilibrium profit per firm and customer on the number of firms for $\gamma = 1$. Numerical results for $m\geq8$ (error bars mark the standard deviation values) are used to fit the dependence $Ar/m^B$; the result is shown with a dashed line. (B) The dependence of the estimated profit exponent on the exponent $\gamma$ of transportation costs. The indicative solid line is $B=1+\gamma/2$. All results for $N=80$.}
\label{fig:sim_s}
\end{figure}

To verify the obtained analytical result, we run numerical simulations with a gradually increasing number of competing firms. Figure~\ref{fig:sim_s}A shows how the average profit per firm and customer decreases with the number of firms $m$. We use weighted least-squares to fit the simulation results obtained for $m\in\{8, 16, 32, 64\}$ (smaller values of $m$ are ignored because the power-law scaling of the profit per firm is expected to hold only for large $m$) with $X(r, m) = Ar / m^B$. The resulting parameter values are $A=0.32\pm0.02$ and $B=1.50\pm0.07$; the value of $B$ is in an excellent agreement with the analytical result $1.50$ contained in Eq.~(\ref{number_of_firms}). By contrast, the value of $A$ is substantially smaller than the analytical value of $1$. In summary, Eq.~(\ref{number_of_firms}) produces a correct scaling of the equilibrium profit per firm with the number of firms, yet it overestimates the profit's absolute value by a factor of three. The reason for this discrepancy is simple. In deriving the analytical result, we assumed that each firm competes only with its closest neighbor. However, each firm is in fact surrounded by a few firms which are all similarly close. The true level of competition is therefore more fierce than we assumed and it is natural to expect that Eq.~(\ref{number_of_firms}) overestimates the average profit of competing firms.

\section{Non-linear transportation costs}
The last question to address is that of a general transportation cost dependence in the form $rd^{\gamma}$ where $r$ is a proportionality term, $d$ is the distance between the customer and the firm and $\gamma>0$ is a distance exponent. By repeating the same progression of steps as described above for $\gamma=1$, we obtain estimates of the exponent $B$ in the power-law relation between the equilibrium profit per firm and the number of competing firms. The obtained results are shown in Figure~\ref{fig:sim_s}B. In the frequently-studied case of quadratic transportation costs, the measured profit scaling exponent is $B=1.98\pm0.06$ (the total profit of all firms thus decays as $1/m$). Overall, the numerical results suggest that $B=1+\gamma/2$ holds in the range $\gamma\in(0,2]$.

However, as can be seen in Figure~\ref{fig:sim_s}B, the initial linear growth of $B$ with $\gamma$ becomes sub-linear for $\gamma\gtrsim2$. One can hypothesize that the change of the dependence of $B$ on $\gamma$ at $\gamma=2$ is due to the special standing of quadratic transportation costs with respect to the equilibrium existence \cite{champsaur1988existence,tabuchi1994two}. To understand why $B$ should not grow with $\gamma$ without bounds, we formulate the following approximate reasoning. We consider the case of large $\gamma$ and introduce the total profit of all firms $T(m)=mX(m)$. When a new firm $l$ is introduced in the system, this has only local consequences because quickly-growing transportation costs---only the firms in direct vicinity are affected by the new competitor. The new total profit $T(m+1)$ can be thus written as
$$
T(m+1)\approx T(m) + \sum_{k\sim l} \Delta X_k + X_l
$$
where $\Delta X_k$ is the induced change of profit of all firms in the vicinity of $l$ and $X_l$ is the profit of the newly added firm. How fast can $T(m)$ decrease with $m$? To see the fastest possible decrease, we assume that $X_l$ is negligible and $\Delta X_k = -X_k(m)$ (the profit of neighboring firms evaporates). Assuming that there are $\varTheta$ neighboring firms in total, we now have $T(m+1)\approx T(m) - \varTheta X(m)$. Since $X(m) = T(m) / m$, we can write $T(m + 1) = T(m) - \varTheta T(m) / m$ which can be converted to the differential equation
$$
T(m+1)-T(m)\approx \mathrm{d}T/\mathrm{d}m = -\varTheta T(m)/m
$$
whose solution is $T(m) = C / m^{\varTheta}$. Since $\varTheta$, the number of neighbors of the newly introduced firm $l$, is a small number, we see that the profit exponent $B$ indeed cannot be arbitrarily large. This explains the saturation of the growth of $B$ with $\gamma$ that is depicted in Figure~\ref{fig:sim_s}B. In any case, values of the transportation cost exponent beyond $\gamma=2$ are little relevant as they are difficult to justify in a realistic setting.

\section{Discussion}
We studied the problem of spatial firm competition with linear transportation costs which, after it initially spurred considerable interest~\cite{hotelling1929stability}, has been much neglected because it does not feature an equilibrium in a case with two firms that are located sufficiently close to each other~\cite{d1979hotelling}. We show here that the problem of equilibrium non-existence does not occur when periodic boundary conditions are considered; the equilibrium then emerges for firms at any distance regardless of the initial conditions. We provide analytical results for the case with two competing firms which compare favorably with extensive numerical simulations of the system. The case of multiple competing firms is considered as well. The main result in this respect is that the total equilibrium profit of all firms decreases as $1/\sqrt{m}$ where $m$ is the number of competing firms. This, as typical for models of spatial firm competition, is in a stark contrast with the basic Bertrand model where the firm profit is zero for any $m\geq2$. Numerical study of general transportation costs that grow as a power of distance shows that the equilibrium profit per firm still has the form $1/m^B$ with the exponent $B$ generally growing with the exponent of the transportation costs.

More generally, our work shows that there is a good reason to study spatial firm competition with linear transportation costs which are arguably more natural and common than quadratic ones. Besides analytically studying the general case of power-law transportation costs, the issue of non-homogeneous customer density also requires attention. Another possibility is to apply the concept of spatial competition on a complex network~\cite{newman2010networks}. While we have assumed here that the firm locations are fixed, the problem of optimal firm location has attracted considerable interest~\cite{gabszewicz1986spatial} and is relevant also here. In addition to freeing the positions of all firms, one can also consider the competition of two firms where each firm can open and locate an arbitrary number of affiliated stores. The equilibrium then emerges when the gain from attracting more customers does not match fixed running costs of new stores.

\begin{acknowledgement}
This work was supported by the EU FET-Open Grant No. 611272 (project Growthcom).
\end{acknowledgement}

\end{document}